**Structure-based inhibitor screening of natural products against NSP15 of SARS-CoV-2 revealed Thymopentin and Oleuropein as potent inhibitors**

**Ramachandran Vijayan**, Samudrala Gourinath*

School of Life Sciences, Jawaharlal Nehru University, New Delhi, India.

*To whom correspondence should be addressed: Dr. Samudrala Gourinath, School of Life Sciences, Jawaharlal Nehru University, Delhi-110067, India.

Email: sgourinath@mail.jnu.ac.in, Tel. No: +91-11-26704513; Fax: +91 -11-26742916/2558

# Structure-based inhibitor screening of natural products against NSP15 of SARS-CoV-2 revealed Thymopentin and Oleuropein as potent inhibitors

**Abstract:** Coronaviruses are enveloped, non-segmented positive-sense RNA viruses that have the largest genome among RNA viruses. The genome contains a large replicase ORF encodes nonstructural proteins (NSPs), structural and accessory genes. NSP15 is a nidoviral RNA uridylate-specific endoribonuclease (NendoU) has C-terminal catalytic domain. The endoribonuclease activity of NSP15 interferes with the innate immune response of the host. Here, we screened Selleckchem Natural product database of compounds against the NSP15, Thymopentin and Oleuropein showed highest binding energies. The binding of these molecules was further validated by Molecular dynamic simulation and found very stable complexes. These drugs might serve as effective counter molecules in the reduction of virulence of this virus. Future validation of both these inhibitors are worth consideration for patients being treated for COVID -19.

**Aims:** NSP15 of SARS-CoV-2 is vital for its life cycle and replication; hence, it is an attractive target for structure-based drug design of anti-SARS drugs. Worldwide research is underway, yet, FDA has not approved any vaccine for its treatment. Infection caused by this virus spreads through human-to-human contact, and has taken a toll of several human lives worldwide. As per the WHO guidelines, lock-down was implemented as a precaution to control the spread of the disease. The outbreak of COVID -19 raised global concerns due to its virulence, and initiated emphasize on the urgent need to find effective drugs for the treatment. **Methods:** The structure-based drug designing approach was utilized to find anti-SARS drugs. **Key findings:** We performed virtual screening of the Selleckchem Natural product database of compounds against the NSP15 of SARS-CoV-2. We have identified Thymopentin (FDA approved drug), Ginsenoside and Oleuropein with high clinical potential against SARS-CoV-2. **Significance:** Further validations are necessary to test the efficacy and safety of these drugs to be used for efficient therapeutics against the COVID -19.

**Keywords:** Drug designing, Virtual screening, Severe Acute Respiratory Syndrome Coronavirus 2 (SARS-CoV-2), Natural products

**Introduction:**

Severe Acute Respiratory Syndrome Coronavirus 2 (SARS-CoV-2) is also known as novel coronavirus of 2019 (COVID -19) [1]. SARS-CoV-2 causes a wide array of respiratory, gastrointestinal, and neurological diseases in humans [2-4]. Coronavirus has the largest genomes of the known RNA viruses, ranging ~30 kb [5], which encodes Structural proteins, S (spike), E (envelope), M (membrane), [6, 7] and Non-structural N (nucleocapsid) proteins [8, 9]. Non-structural protein 15 (NSP15) is an endoribonuclease that cleaves 3' of uridylate through a ribonuclease A (RNase A) [10, 11]. Though it is not required for viral replication, but it mediates evasion of host dsRNA sensors [12, 13]. The structure of SARS-CoV-2 NSP15 protein is very similar to other Coronavirus NSP15 monomers, consisting of mainly three regions: the N-terminal domain, a subsequent middle domain, and a catalytic C-terminal nidoviral RNA uridylate-specific endoribonuclease domain. The active-site residues of NSP15 indicate His262, His277, and Lys317 are conserved across the entire endoribonuclease family and play an important role in enzyme catalysis [13]. The mutant viruses of NSP15 causing apoptotic cell death and significant reduction of replication in macrophages [14, 15]. SARS-CoV-2 has infected more than 14 million people and caused over 532,340 deaths worldwide (https://www.who.int). The rapidly increasing numbers of the infected patients, prompted the World Health Organization to declare a state of global health emergency to coordinate the scientific and medical efforts to expedite the development of a cure for the patients.

NSP15 is important for disease progression and thus, is a potential target for drug and vaccine development. Till date, no targeted treatment molecule has been developed to prevent the COVID-19 infection. Recently *in-silico* studies have reported the possible inhibitors against NSP15 [16-20]. Remedesvir and Favipiravir are the latest approved drugs for the treatment of COVID-19 [21, 22]. However, these drugs are not able to control the novel Coronavirus outbreak and the subsequent pandemic [23]. In order to design specific inhibitors against the Non-structural protein 15 (NSP15), Libraries of Selleckchem Natural products (https://www.selleckchem.com/screening/natural-product-library.html) were chosen for Virtual screening [24-26]. Top 10 compounds were selected based on their binding affinities. Further, Molecular dynamic simulation [25-27] was used to analyze the stability and the inter-molecular interactions between the NSP15 and the lead compounds.

**Materials and methods:**

The crystal structure of the Non-structural protein 15 (NSP15) was retrieved from the RCSB Protein Data Bank (entry code 6W01) and, was used as target for our modeling studies. Starting structure of NSP15 for docking studies was prepared with Protein Preparation Wizard [28]. The process adds hydrogen, neutralizes appropriate amino acid chains, and relieves steric clashes. Also, it performs a series of restrained, partial minimizations on the co-crystallized structure, each of which employs a limited number of minimization steps. It is not intended to minimize the system completely. In our study, the minimization (OPLS 2003 force field) [28] was stopped when RMSD of the non-hydrogen atoms reached 0.30 Å, which is the default specified limit.

Selleckchem Natural product libraries with about 900,000 unique molecules and 1,350,000 drug-like and lead-like screening compounds were chosen for virtual screening. Libraries were prepared by energy minimization of 100 steps with the Ligprep module of Schrodinger [29].

Experimental studies on NSP15 have revealed that His235, His250, and Lys290 [10] are the critical residues for catalysis that are located in the C-terminal domain of the enzyme. Docking studies [30-34] were performed using GOLD [35], and the protein was kept as a rigid molecule, and the number of Genetic Algorithm (GA) runs was set to 10 runs per ligand with the default search algorithm parameters. GOLD score was then used as the final scoring method.

In order to estimate the accuracy of binding affinity of GLIDE [36], MOE docking [31] and Auto-dock [37] were used for cross-docking analysis.

In the Glide-docking [36], the prepared structure was used to generate the receptor grid and no scaling was done for the Van der Waals radii of nonpolar receptor atoms. An enclosing box was used as the docking space, centered on the His235, His250, and Lys290, catalytic triad using the crystallographic position as their reference; the ligand diameter midpoint box was set to the default value (10 Å). Docking experiments were performed using 0.80 to scale the VdW radii of the nonpolar ligand atoms with a charge cutoff of 0.15. Poses were discarded as duplicates if both, the rms deviations in the ligand (all atoms) was less than 0.5 Å and maximum atomic displacement was less than 1.3 Å. At most, 10 poses per ligand were retained. GlideScore XP [36] was used as the scoring method to finalize the screening.

For MOE docking [31], the protein was kept as a rigid entity, and a maximum of 10 conformations for each ligand was taken using the default parameters of MOE with Triangle Matcher placement. The top ranked conformations of NSP15 with the lead-like compounds docked conformations was stored. On the basis of MOE scoring {Generalized-Born Volume Integral/Weighted Surface Area (GBVI/WSA)}, binding free energy calculation in the S field denotes the score. The GBVI/WSA is a scoring function that estimates the free energy of binding of the ligand for a given pose. For all scoring functions, lower scores indicate the more favorable poses. Top ranked docked conformations of the lead-like compounds were selected for further evaluation.

In the AutoDock 4.2 [37], empirical free energy function and the Lamarckian Genetic Algorithm were used for all the docking calculations and the AutoDockTools (ADT) package [28] was used to generate input files and to analyze the results. The NSP15 was set to be rigid and in small molecules all the torsional bonds were kept free during the docking process. The solvent molecules were not considered during the docking process. A cubic grid box of 60 Å size (x62.520, y73.546, z29.278) with a spacing of 0.375 Å and grid maps was created. The ten best solutions based on the docking scores were retained for further investigations. XSCORE [38] has been used to estimate the binding affinity of the molecules. Interactions between the protein and the compounds were calculated using the Discovery Studio program [39]. All Molecular dynamic simulation (MD) were performed with GROMACS [40], using the GROMOS 53A6 force field [40]. Further details of the simulation protocol have been used as described previously [25-27].

**Results and Discussion:**

Superposition of NSP15 with all other available coronavirus homologues reveals structural conservation amongst these proteins. The structure of the SARS-CoV-2 yielded a RMSD of 0.4 Å and 0.9Å with SARS-CoV, and Murine-CoV NSP15, respectively (Figure 1). Comparison of all the 3 structures showed some deviations in the active site residues as marked in Figure 2. These conformational changes at the active site of NSP15 need to be considered carefully during the inhibitor design.

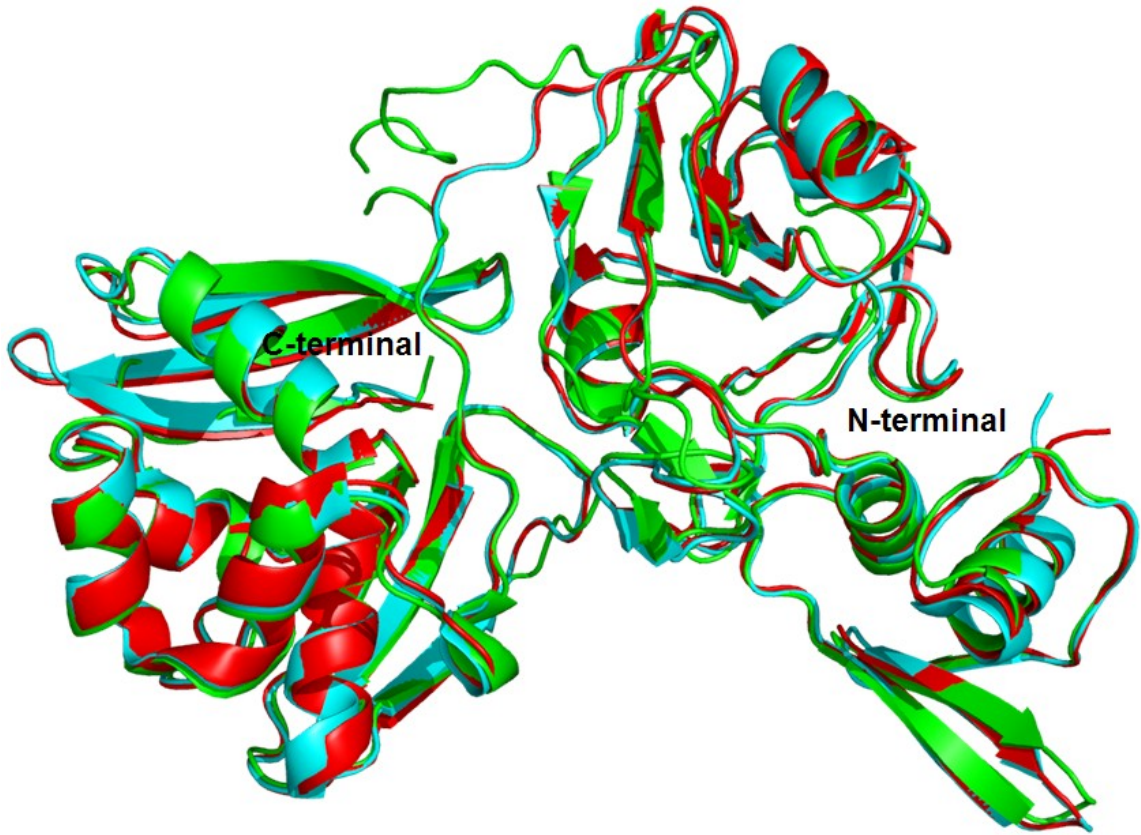

**Figure 1: Superimposition of NSP15 at the active site from various organisms, SARS- CoV-2, SARS-CoV and Murine-CoV showed that the structures are very similar to each other. SARS- CoV-2 depicted in red, SARS-CoV depicted in green, Murine-CoV, depicted in cyan**

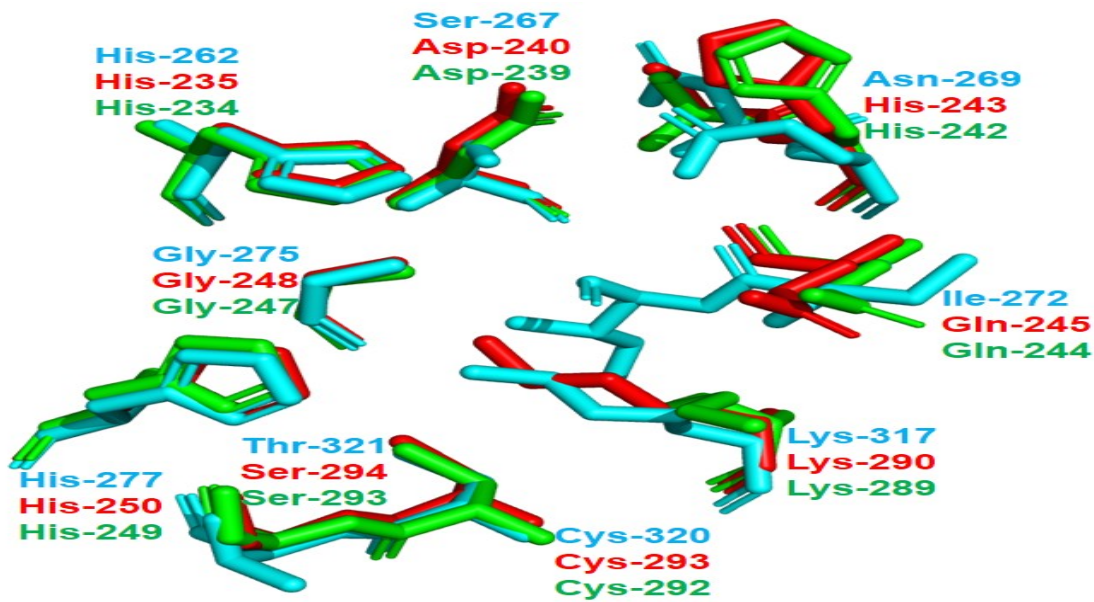

**Figure 2: Residues are shown in the drug-binding region. These conformational changes observed at the active site residues are highlighted. SARS- CoV-2 depicted in red, SARS-CoV depicted in green, Murine-CoV, depicted in cyan**

NSP15 is a known antiviral drug target and its active site constitutes the catalytic residues His 235, His 250, Lys 290. In order to find drugs that could potentially inhibit NSP15, the crystal Structure of NSP15 endoribonuclease from SARS-CoV-2 in complex with citrate at 1.9 Å (PDB ID: 6W01) was used as the target to perform Virtual screening with the Selleckchem Natural Product database containing 28,228 small molecules. His 235, His 250, Lys 290, Thr 341, Tyr 343, and Ser 294 comprises the active site. Among those residues, His 235, His 250, and Lys 290 form the catalytic triad (Figure 2).

GOLD [35] was used to perform virtual screening. Further, multiple docking algorithms using GLIDE [36], Auto-dock [37] and MOE [31] docking ensures the accurate prediction of binding affinities. These docking algorithms are the most commonly used to estimate the binding affinity and binding pose of the inhibitor and the enzyme. The identification of the inhibitors for various targets using these docking and virtual screening algorithms obtained high successful rate based on our previous studies [24-27]. Therefore, we used all of the above-mentioned software to dock the Selleckchem Natural Product database (https://www.selleckchem.com/screening/natural-product-library.html) of 28,228 small molecules against NSP15 in order to evaluate their binding affinities.

Also, we docked the previously reported ligands [16-20, 41, 42], Remedesvir [21, 22] and Lopinavir [21, 22] as reference.

The top-ranked docked molecules were sorted according to their GOLD score [35], GLIDE score [36], Auto-dock free energy of binding [37] and MOE score [31] and XSCORE [38]. These criteria were applied for the higher binding affinity compounds against NSP15 enzyme. Mainly the mode of binding and their potential interactions with the active site residues, and the estimated binding affinities were the major factors considered (Table 2) (Figure 3 & Figure 4).

We have chosen the top 10 compounds (Figure 5) for further analysis since these drug compounds yielded higher binding affinities (Table 1). The top 10 compounds were found to fit tightly into the hydrophobic binding pocket and in close interaction with the catalytic triad. Among these compounds, five compounds show higher binding affinities compared to the reported compounds for NSP15 (Table S1) (Figure S1). The toxicity analysis does not indicate any adverse effects for the top compounds (Table S2), and can be thus taken forward for further studies.

After the docking studies, the Molecular dynamic simulation [25-27] was performed for the top five screened compounds (Thymopentin, Ginsenoside, Oleuropein, Akebia Saponin D and Keampferitrin), to understand the binding stability of the docked complexes. The simulations were performed for a 100 ns to analyze the conformational stability of these complexes.

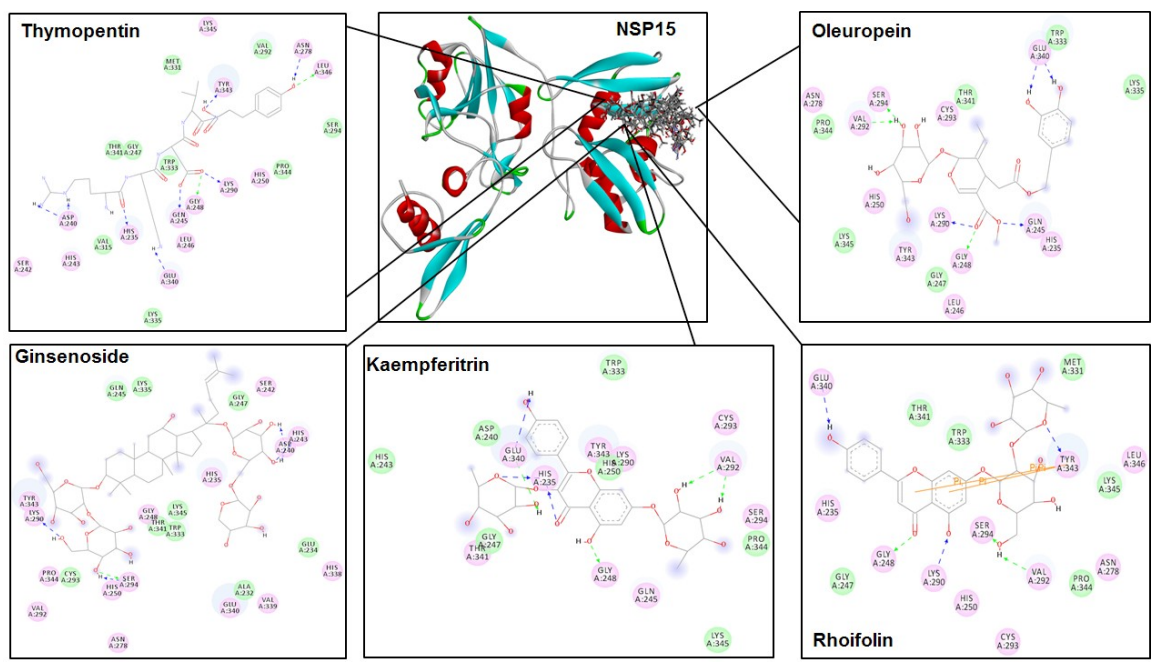

**Figure 3: Modes of binding and its interactions of the top five-inhibitor leads with NSP15**

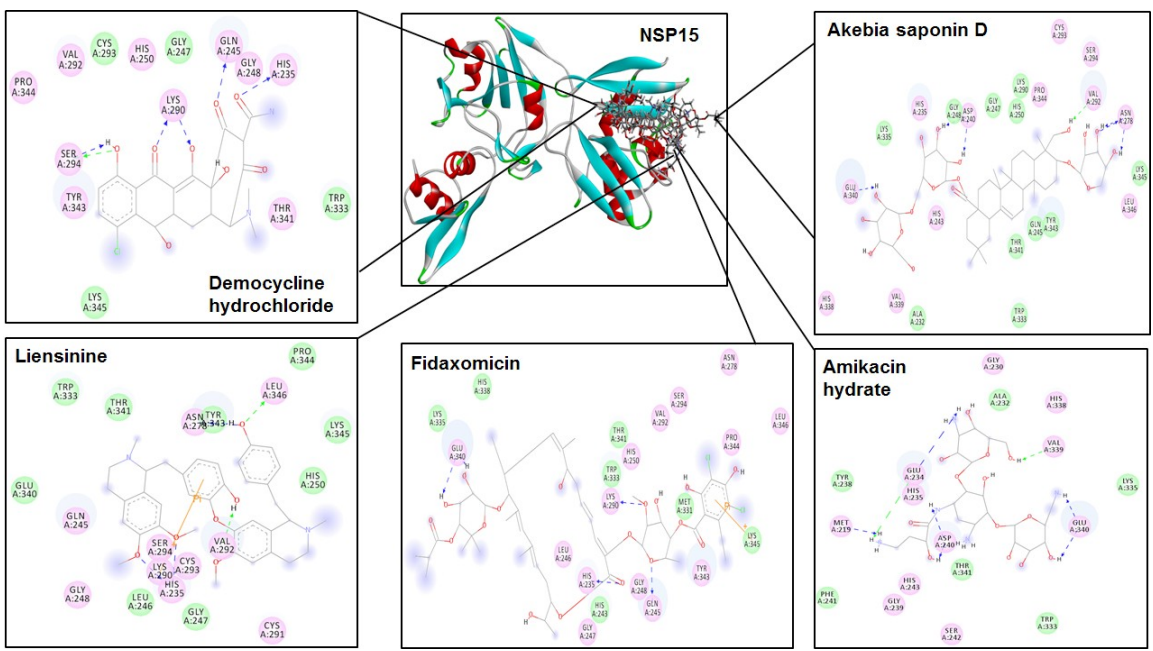

**Figure 4: Modes of binding and its interactions of the top inhibitor leads with NSP15**

| No | Protein | Ligands | GOLD score | GLIDE score kcal/mol | Auto-dock kcal/mol | MOE kcal/mol | X-Score kcal/mol | -logKd |
|---|---|---|---|---|---|---|---|---|
| 1. | NSP15 | Thymopentin | 86.59 | -11.030741 | -9.7 | -12.811 | -9.63 | 8.49 |
| 2. | NSP15 | Oleuropein | 84.69 | -10.960358 | -8.5 | -17.1605 | -8.70 | 7.50 |
| 3. | NSP15 | Ginsenoside | 83.83 | -9.856758 | -8.0 | -14.4559 | -8.92 | 6.54 |
| 4. | NSP15 | Kaempferitrin | 83.32 | -9.17588 | -7.0 | -17.3375 | -9.04 | 7.63 |
| 5. | NSP15 | Rhoifolin | 82.14 | -9.231593 | -6.5 | -12.1433 | -8.43 | 7.18 |
| 6. | NSP15 | Demeclocyclin hydrochloride | 81.71 | -8.553255 | -8.1 | -18.0979 | -8.15 | 6.98 |
| 7. | NSP15 | Akebia Saponin D | 80.17 | -8.173799 | -7.5 | -12.3011 | -8.89 | 7.25 |
| 8. | NSP15 | Liensinine | 79.31 | -7.752487 | -7.7 | -12.4259 | -8.10 | 6.67 |
| 9. | NSP15 | Fidaxomicin | 79.14 | -10.166778 | -7.6 | -15.6436 | -8.58 | 7.10 |
| 10. | NSP15 | Amikacin hydrate | 73.85 | -8.754372 | -7.0 | -12.7334 | -7.80 | 6.12 |

Table 1: Screening of NSP15 with the top-10 inhibitors and their estimated binding affinities

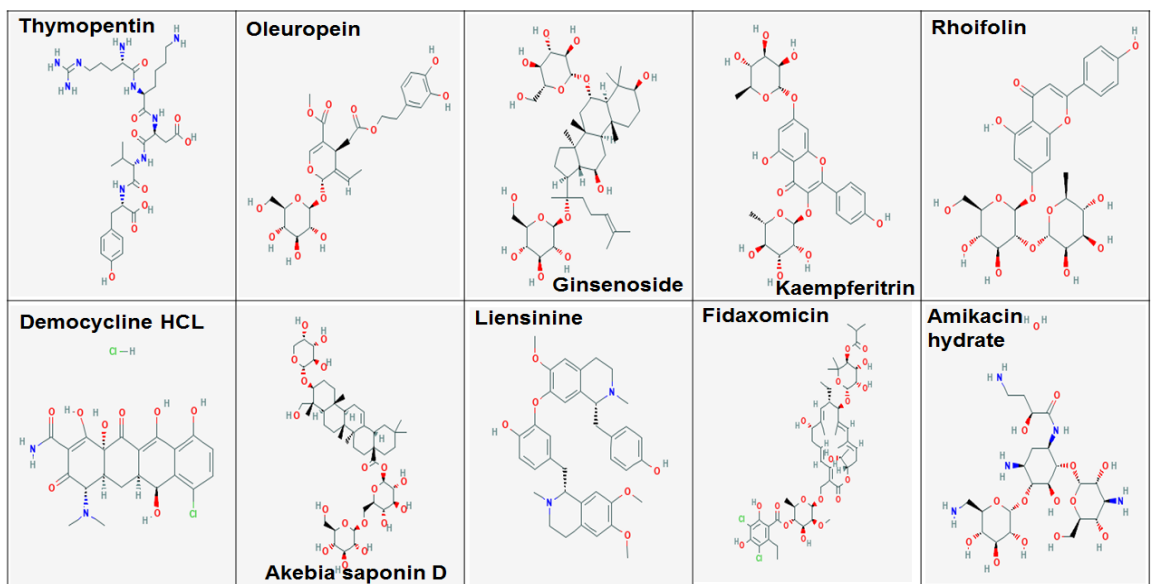

Figure 5: The chemical structure of the top scoring compounds identified from Virtual screening.

Molecular dynamic simulation [25-27] is considered to be a reliable approach with greater insight into the dynamic behavior of proteins and that of the ligand conformations [25-27]. The simulations provide detailed information to better understand the motions of the individual atoms as a function of time and properties of the molecules. We carried out a 100 ns MD simulation to accurately predict the binding stability on the identified compounds against NSP15 as explained in the Methods section. We have studied the root mean square deviation (RMSD), root mean square fluctuation (RMSF), radius of gyration (Rg), solvent accessible surface area (SASA), and hydrogen bonding interactions (NH) between the NSP15 alone as well as in complex with the top five inhibitors. A total of 6 independent simulations were carried out, each with 100 ns simulation time.

According to the RMSD (Root-Mean-Square Deviation) calculation, shown in Figure 5, it can be seen that all the complexes tend to achieve equilibrium after 10 ns and lead to a stable trajectory throughout the simulation. We found that Ginsenoside, Kaempgeritrin and Akebia saponin D complexes with NSP15 tend to reach a higher equilibrium compared to the native, and remained distinguished throughout the simulation, resulting in the RMSD of 0.1 to 0.75 nm. The RMSD trajectories for the Thymopentin and Oleuropein complexes with NSP15 observed a lower equilibrium compared to the native with minor difference in their trajectory that leads to a stable equilibrium through the end of the simulation, suggesting that the complexes stabilized themselves (Figure 6A). The higher RMSD obtained for all the complexes were limited to 0.75 nm that demonstrates the stable trajectories and provided us with an appropriate basis for further investigation. (Figure 6A).

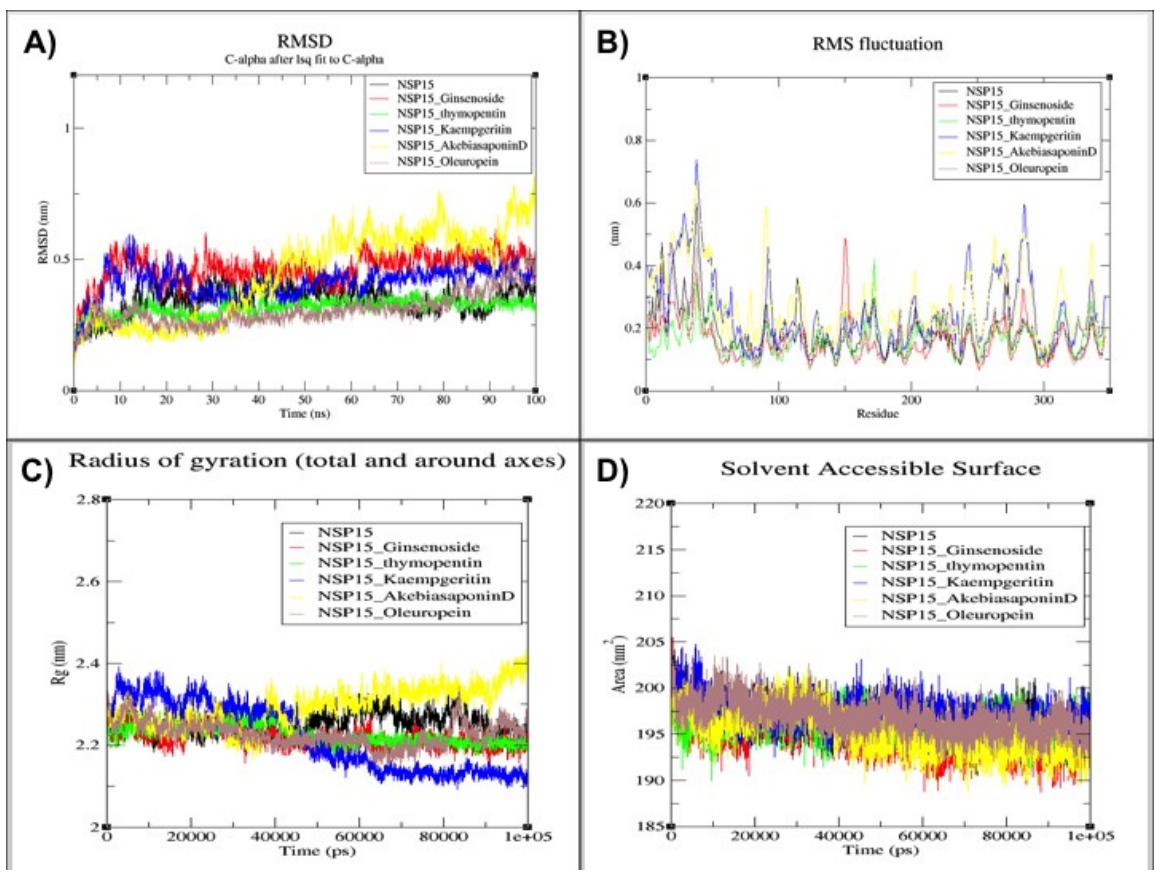

**Figure 6: MD simulation results for NSP15 with ligands. (A) RMSD analysis for the model system NSP15 protein alone and protein in complex with hits compounds (B) Radius of gyration plot for NSP15 and top5 compounds complexes. (C) RMSF plot for NSP15 and top5 hit compounds complexes D) SASA plot for NSP15 and top5 hit compounds complexes**

In order to observe the ligand induced conformational changes of the protein, we calculated the C-RMSF to observe the overall flexibility of the atomic positions in the trajectory for the native and the protein-ligand complexes (Figure 6B). The flexibility of the residues in each of the protein-ligand was analysed by means of RMSF, where the higher RMSF value describes the higher flexibility. According to Figure 5B, Akebia Saponin D and Keampgertrin produced higher fluctuations in the residues positions 40, 90, 110, 200, 250, 275, 290, 310, and 320 during the simulation, while low fluctuations were observed for Ginsenoside, thymopentin, Oleuropein, except at the 150$^{th}$ residue position. Overall, the RMSF result showed that the Ginsenoside, Thymopentin, and Oleuropein complexes were more stable than the Akebia Saponin D and Keampgertrin complexes. (Fig. 6B).

The Rg (radius of gyration) is used to calculate the compactness of a protein (Figure 5C). The radius of gyration is the root mean square distance of a particular atom or group of atoms with its center of mass. The overall NSP15 structure at various time points during the trajectory can be analyzed for the competence, shape, and folding in the Rg plot (. (Figure 5C). Akebia Saponin D-NSP15 complex showed a higher deviation with a Rg score of 2.4 nm. All the other compounds, Ginsenoside, Thymopentin, Oleuropein, and Keampgertrin had an aggregate Rg score of 2.2 nm in a decreasing trend followed by stabilisation after 40 ns towards the end of the simulation (Figure 6C).

We have measured the compactness of the hydrophobic core by analyzing the changes in SASA. As shown in Figure 5D, an increase of the SASA values were observed for Akebia Saponin D and Keampgertrin after 10 ns of simulation time maintained till the end, while the Ginsenoside, Thymopentin, and Oleuropein complexes showed stable conformation throughout the simulations. As higher SASA value leads to an increased exposure, there was loss of hydrophobic contact between the protein-ligand complexes (Figure 6D).

Finally, the number of hydrogen bonds for each of the protein complex was calculated during the simulation. An increase of hydrogen bonds, Van der Waals, and electrostatic interactions were observed for Ginsenoside, Thymopentin, Oleuropein compounds, while Akebia Saponin D and Keampgertrin formed lesser interactions (Fig. 3 & Fig. 4).

The top-ranked compound obtained from the virtual screening is the FDA approved drug, Thymopentin (Table 1). Thymopentin binds to the active site of NSP15 endoribonuclease and may potentially block the replication of virus. Thymopentin (TP5) is a synthetic pentapeptide (Arg-Lys-AspVal-Tyr) belongs to the native thymic hormone thymopoietin [43]. Thymopentin has been suggested for the treatment of autoimmune diseases, and including chronic lymphocytic leukemia [44], rheumatoid arthritis [45], cancer immunodeficiency [46], acquired immunodeficiency syndrome (AIDS) [47], and chronic heart failure [48]. Thymopentin [45] is also known to be an effective immunomodulatory agent and helps to improve immunological condition for patients.

The second compound, Oleuropein showed strong binding with NSP15 endoribonuclease. Oleuropein is a phenylethanoid, mainly found in the olive leaves. Oleuropein suppresses cancer cells by activating the gerosuppressor AMPK by reduction of growth in human primary cells leads to several transcriptomic signatures [49]. Oleuropein helps reducing colonic microflora by Hydroxytyrosol (HT) and is mainly present in the olive leaf and olive oil. Oleuropein is known to be a potent antioxidants observed in nature to date [50].

The third rank compound, Ginsenoside also binds tightly to the catalytic center of NSP15 structure. Ginsenosides are the natural products of steroid glycosides and triterpene saponins. Ginsenosides family consists of the oleanane family and is pentacylic in nature, composed of a five-ring carbon skeleton [51]. Ginsenosides have shown diverse pharmacological and biological properties, such as antitumorogenic, anti-inflammatory, antioxidant, and inhibitor of cell apoptosis.

The top ranking compound Thymopentin [43-48], FDA approved drug is currently available in the market. The other two compounds, Oleuropein [49, 50] and Ginsensoside [51] are also promising ones. Hence, these drugs might help to reduce the virulence of the virus and can be considered for patients being treated for COVID -19.

**Conclusion:**

In this study, structure based virtual screening followed by the validation through Molecular dynamic simulation approaches were carried out to find antiviral leads against NSP15 of SARS-CoV-2. Thymopentin (FDA-approved), Ginsenoside, and Oleuropein were identified as potential inhibitors of NSP15. Molecular docking and MD simulations investigated the binding affinities, mode of binding, stability of binding and their potential interactions. The drug leads identified in this study sheds light on the pandemic infectious disease that currently lacks specific drugs and vaccines. Further investigations will be carried out for these drugs to check their efficiency *in vitro* and *in vivo*.

**Supplementary data**

| No | Protein | Ligands | Reference | MOE score kcal/mol |
|----|---------|---------|-----------|--------------------|
| 1 | NSP15 | Withanoside X | [16] | -6.6530 |
| 2 | NSP15 | Saikosaponins | [17] | -7.8399 |
| 3 | NSP15 | Glycyrrhizic acid | [18, 41, 42] | -7.2673 |
| 4 | NSP15 | Lopinavir | [18, 21, 22] | -5.0902 |
| 5 | NSP15 | Ribavirin | [18] | -3.7347 |
| 6 | NSP15 | Glisoxepide | [19] | -5.0389 |
| 7 | NSP15 | Idarubicin | [19] | -5.5733 |
| 8 | NSP15 | quinadoline B | [20] | -4.0146 |
| 9 | NSP15 | Remedesivir | [21,22] | -5.5850 |

**Table S1: Docking of NSP15 with the reported inhibitor and their estimated binding affinities**

| No | Ligands | Tumori-genicity | Irritant | Drug Likeness | CLogP | CLogS | LE |
|----|---------|-----------------|----------|---------------|-------|-------|-----|
| 1. | Thymopentin | None | None | -2.0657 | -8.3593 | -2.129 | 0.17628 |
| 2. | Oleuropein | None | None | -7.1766 | -0.3124 | -1.768 | 0.22626 |
| 3. | Ginsenoside | None | None | -11.64 | -0.1028 | -4.833 | 0.10914 |
| 4. | Kaempferitrin | None | None | 1.9289 | -0.1371 | -3.387 | 0.20872 |
| 5. | Rhoifolin | None | None | 1.1329 | -2.142 | -2.949 | 0.20874 |
| 6. | Demeclocyclin hydrochloride | None | None | 2.8917 | -4.6648 | -2136 | 0.27149 |
| 7. | Akebia Saponin D | None | None | -12.698 | 0.2283 | -4.85 | 0.12731 |
| 8. | Liensinine | None | None | 4.2826 | 4.0339 | -5.513 | 0.18942 |
| 9. | Fidaxomicin | None | None | -2.7125 | 8.3037 | -7.667 | 0.11386 |
| 10. | Amikacin hydrate | None | None | 2.5019 | -21.186 | 0.233 | 0.21365 |

**Table S2: Various physico-chemical and predicted biological properties of NSP15 and the lead compounds**

**Figure S1: Docking of NSP15 with the reported inhibitor and its interactions**

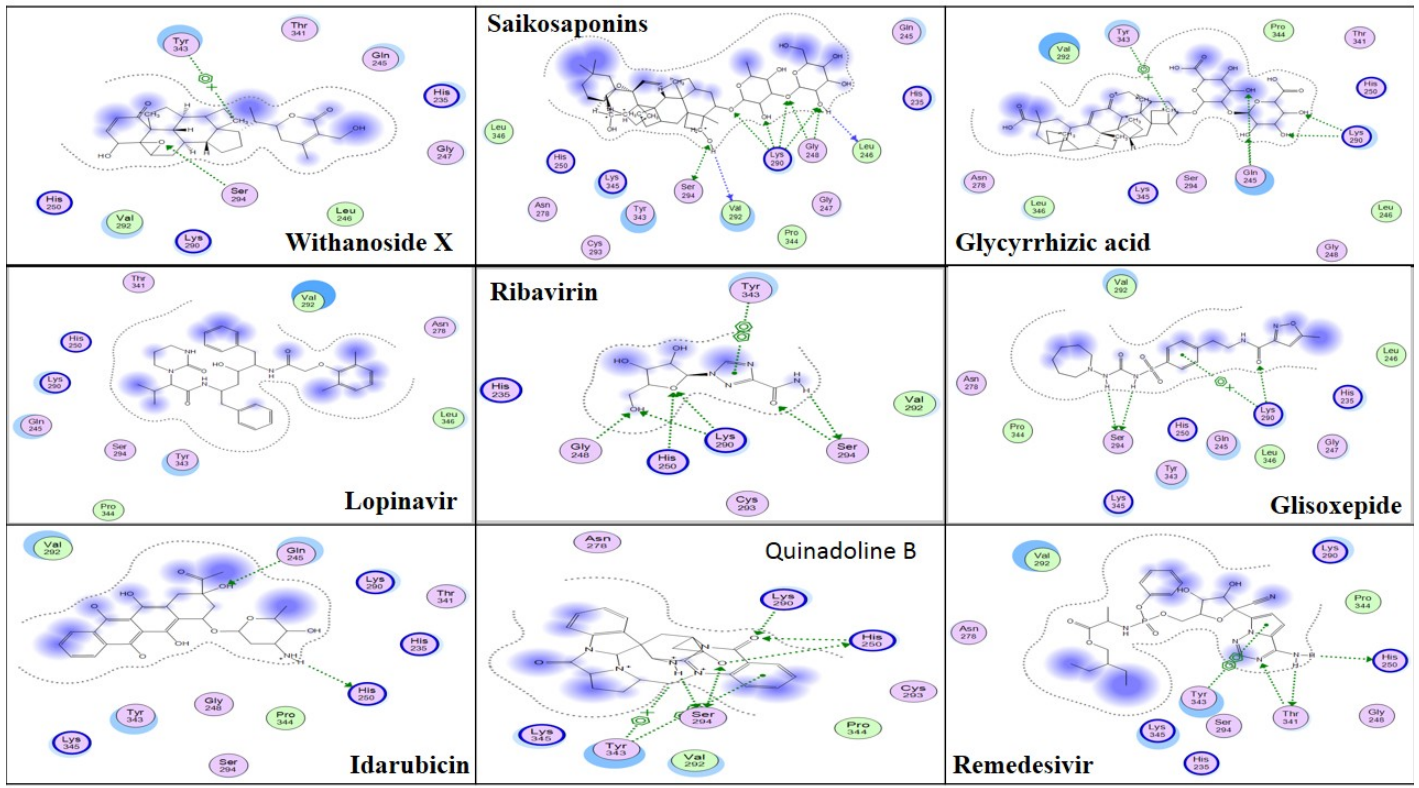